# A generalized Fourier transform of the *P*-quasi distribution function

**Dušan POPOV**
**University Politehnica Timisoara, Department of Physical Foundations of Engineering,**
**Bd. V. Parvan No.2, 300223 Timisoara, Romania**
**dusan_popov@yahoo.co.uk**
**ORCID: https://orcid.org/0000-0003-3631-3247**

***Abstract.***
In the paper we made a generalization of the Fourier transform in the complex space, applicable to the pair of Husimi's and *P*-quasi distributions, in the representation of nonlinear coherent states. Implicitly, our result is a generalization of Mehta's similar result, but which referred only to the canonical coherent states (associated with the one-dimensional harmonic oscillator). Our result is valid for both types of coherent states (defined in the Barut-Girardello, respectively Klauder-Perelomov manner).

## 1. Introduction

As is well known, the coherent states (CSs) of the linear harmonic oscillator (HO-1D) have been discovered by Schrodinger, in 1926 [Schrodinger, 1926], and since then it has always been in the attention of physicists, due to its specific applications, from mathematical physics and solid state physics to cosmology. Moreover, scientific interest and research were not limited to the coherent states and squeezed states of HO-1D, but they were also extended to other quantum systems, making a generalization of the concept of coherent state. Benefiting from a solid foundation of the theory of canonical coherent states (CCSs), through the important works of Glauber [Glauber, 1963] (awarded one half of the 2005 Nobel Prize in Physics "for his contribution to the quantum theory of optical coherence"), Klauder [Klauder, 1963], Sudarshan [Sudarshan, 1963] (missed Nobel Prize several times [Chatterjee et al., 2019]), Perelomov [Perelomov, 1972] and many others, the theory of CSs developed in the sense of their generalization. With the emergence of the generalization process, the canonical coherent states (CCSs), related to HO-1D, were considered to be linear, as opposed to the generalized ones that sound nonlinear. Generalized CSs were thus born, and among its - nonlinear CSs (NCSs), based on the deformation of the bosonic annihilation $a$ and creation $a^+$ canonical operators for HO-1D, as well as the number operator $\hat{n} = a^+ a$. In this sense, the canonical CSs related to HO-1D were considered to be linear.

The generalized NCSs was defined by using the generalized annihilation $A_- = a f(\hat{n})$ and their conjugate, the generalized creation $A_+ = \left(A_-\right)^+ = f^+(\hat{n}) a^+$ operators and then using the usual definitions of CSs, by replacing $a \to A$ and $a^+ \to A^+$ [Matos Filho, 1996], [Man’ko, 1997]. The intensity dependent function $f(\hat{n})$ is called the nonlinearity function and it can be complex or real. Because their phase is irrelevant, one may choose to be real and nonnegative, i.e. $f^+(\hat{n}) = f(\hat{n})$. The normalization function of the most generalized NCSs is a hypergeometric generalized function (HGf) ${}_pF_q(\{a_i\}_1^p ; \{b_j\}_1^q ; |z|^2)$, if the associated Hilbert space is infinite dimensional, that is $n_{\max} \to \infty$, or a hypergeometric generalized polynomial (HGp), ${}_p^{n_{\max}}F_q(\{a_i\}_1^p ; \{b_j\}_1^q ; |z|^2)$, if Hilbert space is finite dimensional, i.e. with $n_{\max} < \infty$. These kinds of states were firstly introduced by Appl and Schiller [Appl, Schiller, 2004] and applied to the mixed (thermal) states of the pseudoharmonic oscillator in one of our previous papers [Popov, EJTP, 2006].

From the very beginning of the crystallization of the CSs theory, the following problem was posed: if, knowing the diagonal elements of the density operator in the CSs representation (that is, knowing Husimi's distribution function $Q(z) \equiv < z | \rho | z >$), the expression of afferent *P*-quasi distribution function can be obtained. For canonical CSs, the problem was first solved in 1967 by Mehta [Mehta, 1967], the respective relationship being sometimes called "Mehta's formula", so that it can then be deduced by other methods. According to our knowledge, in the specialized literature, this problem has not been solved for NCSs. That is why this is the purpose of the present paper.

The paper is organized as following: In Sec. 2 we present the basic notions regarding NCSs and the rules of the diagonal operator ordering technique (DOOT), which will be necessary for the approach in the following sections. Sec. 3 is dedicated to reviewing the methods of deducing Mehta's formula. Sec. 4 deals with the generalization of Mehta's formula for a certain NCSs state, with rigorous checks. In Sec. 5 we illustrate the veracity of the theoretically deduced considerations in the previous section, examining an example, namely the pseudoharmonic oscillator (PHO). Sec. 6, the last one, presents some theoretical considerations regarding the generalization of Mehta's formula for Barut-Girardello and Klauder-Perelomov nonlinear coherent states associated with a quantum system.

## 2. Some basics elements of generalized coherent states using the DOOT

We consider a pair of annihilation and creation operators $A_-$ and $A_+$, whose action on the Fock vectors $| n >, \;\; n = 0,1,2,...$ are

$$A_- | n > = \sqrt{e(n)} \, | n-1 > \quad , \quad A_+ | n > = \sqrt{e(n+1)} \, | n+1 > \quad , \quad A_+ A_- | n > = e(n) | n > \qquad (2.1)$$

Generally, the operators $A_-$ and $A_+$ are not commutable, but, as we will see below, they obey the rules of the diagonal operator ordering technique (DOOT).

Applying *n*-times the creation / annihilation operators on the vacuum states $| 0 >$, respectively $< 0 |$, we obtain

$$| n > = \frac{1}{\sqrt{\rho(n)}} \left( A_+ \right)^n | 0 > \, , \; < n | = \frac{1}{\sqrt{\rho(n)}} < 0 | \left( A_- \right)^n , \; | n >< n | = \frac{1}{\rho(n)} \left( A_+ \right)^n | 0 >< 0 | \left( A_- \right)^n \qquad (2.2)$$

Here appear a real and positive *structure function* $\rho(n)$ , defined as

$$\rho(n) \equiv \prod_{s=1}^{n} e(s) = \prod_{s=1}^{n} s\left[f(s)\right]^2 = n!\left(\left[f(n)\right]!\right)^2 , \quad \left[f(n)\right]! = f(1)f(2)...f(n) , \quad f(0)=1 . \qquad (2.3)$$

where $f(n)$ is a *non-linearity function* depending on the main quantum number $n$, the eigenvalue of the number operator $\hat{n}\,|\,n>=n\,|\,n>$. This will determine the non-linearity degree of the coherent states.

We will choose the non-linearity function of a special manner:

$$f(n) = \sqrt{\frac{\prod_{j=1}^{q}\left(b_j - 1 + n\right)}{\prod_{i=1}^{p}\left(a_i - 1 + n\right)}} \quad , \qquad e(n) = n\frac{\prod_{j=1}^{q}\left(b_j - 1 + n\right)}{\prod_{i=1}^{p}\left(a_i - 1 + n\right)} \qquad (2.4)$$

and consequently the structure function is

$$\rho(n) = n!\prod_{s=1}^{n}\left[f(s)\right]^2 = n!\left(\left[f(n)!\right]\right)^2 = n!\frac{\prod_{j=1}^{q}\left(b_j\right)_n}{\prod_{i=1}^{p}\left(a_i\right)_n} \qquad (2.5)$$

In a previous paper [Popov, Popov, 2015] we have introduced a new operator ordering technique, the *diagonal operator ordering technique (DOOT)*, as a generalization applicable to the generalized coherent states (CSs) of a similar technique for canonical CSs – the *integration within an ordered product (IWOP)*, introduced by Hong-yi Fan (see, e.g. [Fan, 1999] and references therein). But, the IWOP is applicable *only for Bose operators*, referring to the CSs of the HO-1D. In this sense, the IWOP appear as a particular case of DOOT.

The main rules of the DOOT are: a) the order of operators $A_-$ and $A_+$ can be permuted inside the symbol # #, i. e. $\#(A_-)^n(A_+)^n\# = \#(A_+)^n(A_-)^n\# = \#(A_+A_-)^n\#$, so that finally we obtain a function of normally ordered operator product $\#f\left(A_+A_-\right)\#$; b) inside the symbol # # we can perform all algebraic operations, according to the usual rules; c) the operators $A_-$ and $A_+$ can be treated as simple *c*-numbers; d) the vacuum state projector $|\,0><0\,|$, in the frame of DOOT, has the following normal ordered form:

$$|\,0><0\,| = \frac{1}{\sum_{n=0}^{\infty}\frac{1}{\rho(n)}\#\left(A_+A_-\right)^n\#} = \frac{1}{\#\,{}_pF_q(\{a\}_1^p\,;\,\{b\}_1^q\,;A_+A_-)\#} \tag{2.6}$$

where in the denominator appear the generalized hypergeometric function which depend on the normally ordered operator product $\#A_+A_-\#$ as "argument". The deduction of the above equality is simple: starting from the completeness relation for the Fock vectors $\sum_{n=0}^{\infty}|\,n><n\,|=1$ and using the results of the multiple actions of operators $A_-$ and $A_+$ on the fundamental (vacuum) state vectors, the above relation is reached.

$$\begin{aligned} 1 &= \sum_n |\,n><n\,| = \sum_n \frac{1}{\rho(n)}\left(A_+\right)^n |\,0><0\,|\left(A_-\right)^n = \\ &= |\,0><0\,|\sum_n \frac{1}{\rho(n)}\#\left(A_+A_-\right)^n\# = |\,0><0\,|\#\,{}_pF_q(\{a\}_1^p\,;\,\{b\}_1^q\,;A_+A_-)\# \end{aligned} \tag{2.7}$$

The choice of the above expression for the nonlinearity function $f(n)$ is closely related to the definition of generalized hypergeometric function ${}_pF_q(\{a\}_1^p\,;\,\{b\}_1^q\,;x)$:

$${}_pF_q(\{a\}_1^p\,;\,\{b\}_1^q\,;x) = \sum_{n=0}^{\infty}\frac{\prod_{i=1}^{p}(a_i)_n}{\prod_{j=1}^{q}(b_j)_n}\frac{x^n}{n!} \equiv \sum_{n=0}^{\infty}\frac{1}{\rho(n)}x^n \tag{2.8}$$

where $p$ and $q$ are positive integers, $a_i$ and $b_j$ real numbers.

To simplify the writing of the formulas, *where confusion will not be created*, we will use the following abbreviated notations:

$$\begin{aligned} &\{a_i\}_{i=1}^p = \{a_1\,,\,a_2\,,...,a_p\} \equiv \boldsymbol{a}\;\;,\;\; \{\pm a_i \mp 1\}_{i=1}^p \equiv \pm\boldsymbol{a}\mp\boldsymbol{1}\,, \\ &\{b_j\}_{j=1}^q = \{b_1\,,\,b_2\,,...,b_q\} \equiv \boldsymbol{b}\;,\;\; \{\pm b_j \mp 1\}_{j=1}^q \equiv \pm\boldsymbol{b}\mp\boldsymbol{1}\,, \\ &{}_pF_q(\{a\}_1^p\,;\,\{b\}_1^q\,;x) \equiv {}_pF_q(\boldsymbol{a}\,;\,\boldsymbol{b}\,;x)\;,\;\; {}_q\tilde{F}_p(\{b\}_1^q\,;\,\{a\}_1^p\,;x) \equiv {}_q\tilde{F}_p(\boldsymbol{b}\,;\,\boldsymbol{a}\,;x) \\ &G_{p,q+1}^{1,p}\left(-x\left|\begin{matrix}\{1-a_i\}_1^p\,; & / \\ 0\;; & \{1-b_j\}_1^q\end{matrix}\right.\right) \equiv G_{p,q+1}^{1,p}\left(-x\left|\begin{matrix}\boldsymbol{1}-\boldsymbol{a}\,; & / \\ 0\;; & \boldsymbol{1}-\boldsymbol{b}\end{matrix}\right.\right)\;, \\ &G_{p,\,q+1}^{q+1,0}\left(x\left|\begin{matrix} /\;; & \{a_i-1\}_1^p \\ 0\,,\{b_j-1\}_1^q\;; & / \end{matrix}\right.\right) \equiv G_{p,\,q+1}^{q+1,0}\left(x\left|\begin{matrix} /\;; & \boldsymbol{a}-\boldsymbol{1} \\ 0\,,\,\boldsymbol{b}-\boldsymbol{1}\;; & / \end{matrix}\right.\right) \end{aligned} \tag{2.9}$$

and similarly for other situations. Only for the situations when the sequences of numbers $\{a_i\}_{i=1}^{p}$ and $\{b_j\}_{j=1}^{q}$ *are broken*, we will write them explicitly. The Pochhammer symbols for $n = 1, 2, 3, \ldots$ are:

$$(a)_n = \frac{\Gamma(a+n)}{\Gamma(a)} = a(a+1)\ldots(a+k-1), (-a)_n = (-1)^n \frac{\Gamma(a+1)}{\Gamma(a+1-n)}, \ (0)_0 = 1, \ (0)_n = 0, \ (a)_0 = 1 \quad (2.10)$$

The generalized hypergeometric function can be represented through Meijer G-function [Mathai, 1973]:

$$ {}_pF_q(\boldsymbol{a}; \boldsymbol{b}; x) = \Gamma(b/a)\, G^{1,p}_{p,q+1}\left(-x \middle| \begin{matrix} \boldsymbol{1-a}\ ; & / \\ 0\ ; & \boldsymbol{1-b} \end{matrix}\right) ; \quad \Gamma(b/a) \equiv \frac{\prod_{j=1}^{q} \Gamma(b_j)}{\prod_{i=1}^{p} \Gamma(a_i)} \quad (2.11)$$

Depending on the definition manner, there are three types of generalized coherent states, sometimes also called nonlinear coherent states (NCSs), to distinguish them from canonical coherent states, considered linear states:

1.) The Barut-Girardello coherent states (BG-CSs), which are eigenstates of the annihilation operator $A_-$ [Barut, 1971].

2.) The Klauder-Perelomov coherent states (KP-CSs), defined through group-theoretical approach [Perelomov, 1972].

3.) The Gazeau-Klauder coherent states (GK-CSs), which are nonspreading and temporally stable, directly related to the Hamiltonian of the examined system [Gazeau, Klauder, 1999]. In [Popov, Negrea, 2022] we showed that GK-CSs states can be obtained from BG-CSs, following three steps.

Regardless of its type (definition), the expansion according to the Fock vectors of the NCSs is

$$| z > = \frac{1}{\sqrt{\mathcal{N}(|z|^2)}} \sum_{n=0}^{\infty} \frac{z^n}{\sqrt{\rho(n)}} | n > = \frac{1}{\sqrt{\mathcal{N}(|z|^2)}} \mathcal{N}(zA_+) | 0 > \quad (2.12)$$

with $z = |z| \exp(\mathrm{i}\varphi)$ the complex variable labeling the NCSs, $0 \le |z| \le R_c \le \infty$, $0 \le \varphi \le 2\pi$, and the normalization function $\mathcal{N}\left(|z|^2\right)$ which is obtained from the normalization condition $< z \,|\, z >= 1$ :

$$\mathcal{N}\left(|z|^2\right) = \sum_{n=0}^{\infty} \frac{1}{\rho(n)} \left(|z|^2\right)^n \quad . \tag{2.13}$$

This function must fulfill the condition $0 < \mathcal{N}\left(|z|^2\right) < \infty$ i.e., the convergence radius $R_c$ of the series must fulfill the inequality: $0 < |z| < R_c = \lim\limits_{n\to\infty} \frac{\rho(n)}{\rho(n+1)}$.

On the other hand, all NCSs must fulfill the following minimal conditions generically called “the Klauder’s prescriptions” [Klauder, 1963]:

(I). NCSs are normalized but non-orthogonal:

$$< z \,|\, z' >= \frac{\mathcal{N}\left(z^* z'\right)}{\sqrt{\mathcal{N}\left(|z|^2\right)}\sqrt{\mathcal{N}\left(|z'|^2\right)}} = \begin{cases} 1, & z = z' \\ \neq 0, & z \neq z' \end{cases} \tag{2.14}$$

and form an overcomplete set.

(II). Continuity in the label variable $z$ of the NCSs:

$$\lim_{z\to z'} \| z - z' \| = \lim_{\substack{r\to r' \\ \varphi\to\varphi'}} \sqrt{r^2 + r'^2 - 2rr'\cos(\varphi - \varphi')} = 0 \tag{2.15}$$

(III). NCSs must satisfy the resolution of the identity, i.e. close a resolution of the identity operator

$$\int d\mu(z) |\, z >< z \,| = \hat{I} = \sum_{n=0}^{\infty} |\, n >< n \,| \tag{2.16}$$

where the integration measure is $d\mu(z) = \frac{d^2 z}{\pi} = \frac{d\varphi}{2\pi} d\left(|z|^2\right) h\left(|z|^2\right)$ and its weight function $h\left(|z|^2\right)$ which must be found for each individual case.

To be satisfied the completeness relation $\sum_{n=0}^{\infty} |\, n >< n \,| = 1$, the angular integration must be

$$\int_0^{2\pi} \frac{d\varphi}{2\pi} \left(z^*\right)^n z^{n'} = \left(|z|^2\right)^n \delta_{n\,n'} \tag{2.17}$$

Performing the function change $h_{red}\left(|z|^2\right)=\dfrac{h\left(|z|^2\right)}{\mathcal{N}\left(|z|^2\right)}$, as well as the exponent change $n=s-1$, we have to solve the following moment problem in variable $|z|^2$:

$$\int_0^{R_c} d\left(|z|^2\right) h_{red}\left(|z|^2\right)\left(|z|^2\right)^{s-1}=\rho(s-1)=(s-1)!\prod_{m=1}^{s-1}\left[f(m)\right]^2 \tag{2.18}$$

The value of convergence radius $R_c$ (which is calculable using one of the convergence criteria of the power series, for example the ratio criterion) determines the type of moment problem [Klauder et al., 2001]:

$$R_c=\lim_{n\to\infty}\frac{\rho(n)}{\rho(n+1)}=\begin{cases}\infty, & \text{Stieltjes moment problem} (=\text{SM})\\ <\infty, & \text{Hausdorff moment problem} (=\text{HM})\end{cases} \tag{2.19}$$

This is determined by the values of the integer indices $p$ and $q$:

$$R_c=\lim_{n\to\infty}\frac{\rho(n)}{\rho(n+1)}=\lim_{n\to\infty}n^{p-q-1}\frac{1}{1+\frac{1}{n}}\frac{\prod_{i=1}^{p}\left(\frac{a_i}{n}+1\right)}{\prod_{j=1}^{q}\left(\frac{b_j}{n}+1\right)}=\lim_{n\to\infty}n^{p-q-1}=\begin{cases}\infty, & \text{if} \quad p-q-1>0\\ 1, & \text{if} \quad p-q-1=0\\ 0, & \text{if} \quad p-q-1<0\end{cases} \tag{2.20}$$

In a compact and unitary expression, the moment problem is written in matrix form as [Roknizadeh, Tavassoly, 2004]

$$\int_0^{\infty} d\left(|z|^2\right)\begin{pmatrix}\tilde{h}\left(|z|^2\right)\\ \\ H\left(R_c-|z|^2\right)\tilde{h}\left(|z|^2\right)\end{pmatrix}\left(|z|^2\right)^{s-1}=\rho(s-1) \qquad \leftrightarrow \qquad \begin{pmatrix}SM\\ \\ HM\end{pmatrix} \tag{2.21}$$

Here $H\left(R_c-|z|^2\right)$ is the Heaviside step function

$$H\left(R_c-|z|^2\right):=\begin{cases}0, & |z|^2>R_c\\ 1, & |z|^2\le R_c\end{cases} \tag{2.22}$$

Consequently, the solution $\tilde{h}\left(|z|^2\right)$ is proportional to a Meijer G-function and its concrete expression depends on the type of coherent state. This is a weight function and must necessarily be positive.

$$\tilde{h}\left(|z|^2\right)=C(\boldsymbol{a},\boldsymbol{b})\; G_{p,q}^{m,n}\left(|z|^2\,\middle|\,\begin{matrix}\{a_i\}_1^n\;; & \{a_i\}_{n+1}^p\\ \{b_j\}_1^m\;; & \{b_j\}_{m+1}^q\end{matrix}\right) \tag{2.23}$$

where $C(\boldsymbol{a},\boldsymbol{b})$ is a constant that depends on the internal structure of the nonlinearity function $f(n)$.

Then, the integration measure becomes:

$$d\mu(z) = C(\boldsymbol{a},\boldsymbol{b})\frac{d\varphi}{2\pi}d\left(|z|^2\right)\mathcal{N}\left(|z|^2\right)G_{p,q}^{m,n}\left(|z|^2 \left| \begin{array}{ccc} \{a_i\}_1^n & ; & \{a_i\}_{n+1}^p \\ \{b_j\}_1^m & ; & \{b_j\}_{m+1}^q \end{array} \right.\right) \tag{2.24}$$

According to the DOOT rules, the projector of NCSs is

$$|z\rangle\langle z| = \frac{1}{\mathcal{N}\left(|z|^2\right)}\#\mathcal{N}\left(zA_+\right)|0\rangle\langle 0|\mathcal{N}\left(z^*A_-\right)\# = \frac{1}{\mathcal{N}\left(|z|^2\right)}\#\frac{\mathcal{N}\left(zA_+\right)\mathcal{N}\left(z^*A_-\right)}{\mathcal{N}\left(A_+A_-\right)}\# \tag{2.25}$$

As a result, the resolution of identity is fulfilled if the following integral can be solved

$$\int\frac{d^2z}{\pi}G_{p,q}^{m,n}\left(|z|^2 \left| \begin{array}{ccc} \{a_i\}_1^n & ; & \{a_i\}_{n+1}^p \\ \{b_j\}_1^m & ; & \{b_j\}_{m+1}^q \end{array} \right.\right)\#\mathcal{N}\left(zA_+\right)\mathcal{N}\left(z^*A_-\right)\# = \frac{1}{C(\boldsymbol{a},\boldsymbol{b})}\#\mathcal{N}\left(A_+A_-\right)\# \tag{2.26}$$

This is a new generalized integral in which the operators are regarded as simple c-numbers, which can be replaced by some constants.

The angular integral leads to

$$\int_0^{2\pi}\frac{d\varphi}{2\pi}\#\mathcal{N}\left(zA_+\right)\mathcal{N}\left(z^*A_-\right)\# = \sum_{n=0}^{\infty}\frac{\#\left(A_+A_-\right)^n\#}{[\rho(n)]^2}\left(|z|^2\right)^n \tag{2.27}$$

and so

$$\sum_{n=0}^{\infty}\frac{\#\left(A_+A_-\right)^n\#}{[\rho(n)]^2}\int_0^{\infty}d\left(|z|^2\right)\left(|z|^2\right)^n G_{p,q}^{m,n}\left(|z|^2 \left| \begin{array}{ccc} \{a_i\}_1^n & ; & \{a_i\}_{n+1}^p \\ \{b_j\}_1^m & ; & \{b_j\}_{m+1}^q \end{array} \right.\right) = \frac{1}{C(\boldsymbol{a},\boldsymbol{b})}\#\mathcal{N}\left(A_+A_-\right)\# \tag{2.28}$$

To have equality, the integral must be

$$\int_0^{\infty}d\left(|z|^2\right)\left(|z|^2\right)^n G_{p,q}^{m,n}\left(|z|^2 \left| \begin{array}{ccc} \{a_i\}_1^n & ; & \{a_i\}_{n+1}^p \\ \{b_j\}_1^m & ; & \{b_j\}_{m+1}^q \end{array} \right.\right) = \frac{1}{C(\boldsymbol{a},\boldsymbol{b})}\rho(n) \tag{2.29}$$

In the case of the *mixed states*, particularly the *thermal states*, the approach cannot be performed for the general case, due to the different expressions of the corresponding energy eigenvalues. As a characteristic example of mixed states we will refer to the thermal states of a quantum canonical ensemble at equilibrium temperature $T = 1/(k_B\,\beta_T)$, with energy eigenvalues $E_n$, whose equilibrium density operator $\hat{\rho}$ is

$$\hat{\rho} = \frac{1}{Z(\beta_T)} \sum_n e^{-\beta_T E_n} \, | n >< n | \tag{2.30}$$

and also with the partition function $Z(\beta)$ of the system

$$Z(\beta_T) = \sum_n e^{-\beta_T E_n} \tag{2.31}$$

In the NCSs representation, the ensemble is characterized by the diagonal element of the density operator, i.e. by the *Q*- distribution function (Husimi's function)

$$Q(|z|^2) \equiv < z | \hat{\rho} | z > = \frac{1}{\mathcal{N}\left(|z|^2\right)} \sum_n e^{-\beta_T E_n} \frac{(|z|^2)^n}{\rho(n)} \tag{2.32}$$

as well as the *P*- quasi distribution function. This distribution is also called the *Glauber-Sudarshan (GS) distribution* which yields a diagonal expansion in terms of CSs, which play the role of the weight function in the expansion of density operator $\hat{\rho}$ with respect to the NCSs projector $| z >< z |$ [Glauber, 1963], [Sudarshan, 1963]:

$$\hat{\rho} = \int d\mu(z) \, P(|z|^2) | z >< z | \tag{2.33}$$

Using the DOOT rules, the density operator becomes

$$\hat{\rho} = \frac{1}{Z(\beta_T)} \# \frac{1}{\mathcal{N}(A_+ A_-)} \sum_n e^{-\beta_T E_n} \frac{(A_+ A_-)^n}{\rho(n)} \# \tag{2.34}$$

with their diagonal expansion

$$\hat{\rho} = \frac{1}{Z(\beta_T)} \# \frac{C(\boldsymbol{a}, \boldsymbol{b})}{\mathcal{N}(A_+ A_-)} \int_0^\infty d(|z|^2) \, G_{p,q}^{m,n}\left( |z|^2 \left| \begin{array}{ccc} \{a_i\}_1^n & ; & \{a_i\}_{n+1}^p \\ \{b_j\}_1^m & ; & \{b_j\}_{m+1}^q \end{array} \right. \right) P(|z|^2) \int_0^{2\pi} \frac{d\varphi}{2\pi} \mathcal{N}(zA_+) \mathcal{N}(z^* A_-) \# \tag{2.35}$$

so that we have

$$\hat{\rho} = \frac{1}{Z(\beta_T)} \# \frac{C(\boldsymbol{a}, \boldsymbol{b})}{\mathcal{N}(A_+ A_-)} \# \sum_n \frac{\#(A_+ A_-)^n \#}{[\rho(n)]^2} \int_0^\infty d(|z|^2) G_{p,q}^{m,n}\left( |z|^2 \left| \begin{array}{ccc} \{a_i\}_1^n & ; & \{a_i\}_{n+1}^p \\ \{b_j\}_1^m & ; & \{b_j\}_{m+1}^q \end{array} \right. \right) P(|z|^2) \left(|z|^2\right)^n \tag{2.36}$$

It follows that the integral must be

$$\int_0^\infty d(|z|^2) G_{p,q}^{m,n}\left( |z|^2 \left| \begin{array}{ccc} \{a_i\}_1^n & ; & \{a_i\}_{n+1}^p \\ \{b_j\}_1^m & ; & \{b_j\}_{m+1}^q \end{array} \right. \right) P(|z|^2) \left(|z|^2\right)^n = \frac{1}{Z(\beta_T)} e^{-\beta_T E_n} \frac{\rho(n)}{C(\boldsymbol{a}, \boldsymbol{b})} \tag{2.37}$$

Furthermore, the concrete expressions for $\hat{\rho}$, respectively the *Q*- and *P*- functions are different, due to the different expression for $E_n$. In the present paper we focused only on the quantum systems with *linear energy spectrum*

$$E_n = \hbar\omega\, e(n) = \hbar\omega\left(n + e_0\right) \tag{2.38}$$

For such systems, the density operator and partition function are

$$\hat{\rho} = \frac{e^{-\beta_T \hbar\omega\, e_0}}{Z(\beta_T)} \sum_{n=0}^{\infty} \left(e^{-\beta_T \hbar\omega}\right)^n |\, n><n\,| = \frac{1}{\bar{n}+1} \sum_{n=0}^{\infty} \left(\frac{\bar{n}}{\bar{n}+1}\right)^n |\, n><n\,| \quad , \tag{2.39}$$

where we used the Bose-Einstein distribution function

$$\bar{n} = \frac{1}{e^{\beta_T \hbar\omega} - 1} \tag{2.40}$$

For other more complicated energy spectra, specific methods must be applied. For example, for a quadratic spectrum (characteristic, e.g. of the Morse oscillator) a specific ansatz is applied [Popov, 2003].

Following a traditional way (one can even say "classical"), let's build the NCSs with the operators $A_-$ and $A_+$, in the spirit of the founders of this concept (see, e.g. [Glauber, 1963], [Sudarshan, 1963], [Barut, 1971], [Perelomov, 1972], [Gilmore, 1974], [Klauder, Skagerstam, 1985], and not only).

Using a nonlinearity function of type (2.4), we can construct NCSs of type Barut-Girardello, respectively Klauder-Perelomov. In order not to burden the writing of the formulas and to distinguish the two types of NCSs, for NCSs and related variable and quantities of the Klauder-Perelomov type we will use the "tilda" sign. At the same time, we will write the variable and the quantities that refer to NCSs of the Barut-Girardello type without this sign.

**Barut-Girradello nonlinear coherent states** (BG-NCSs) are defined as eigenvectors of the nonlinear annihilation operator $A_- = a\, f(n)$ [Barut, 1971], [Roy, Roy, 2000]:

$$A_- \,|\, z> = z\,|\, z> \tag{2.41}$$

These can be expanded according to the Fock vectors as

$$|\, z> = \frac{1}{\sqrt{{}_pF_q(\boldsymbol{a}\,;\,\boldsymbol{b}\,;|\,z\,|^2)}} \sum_{n=0}^{\infty} \frac{z^n}{\sqrt{\rho(n)}} |\, n> = \frac{1}{\sqrt{{}_pF_q(\boldsymbol{a}\,;\,\boldsymbol{b};|\,z\,|^2)}}\, {}_pF_q(\boldsymbol{a}\,;\,\boldsymbol{b};|\,zA_+)|\,0> \tag{2.42}$$

The NCSs projector is then

$$|z><z| = \frac{1}{{}_pF_q(\boldsymbol{a};\boldsymbol{b};|z|^2)} \# \frac{{}_pF_q(\boldsymbol{a};\boldsymbol{b};|zA_+)\ {}_pF_q(\boldsymbol{a};\boldsymbol{b};|z^*A_-)}{{}_pF_q(\boldsymbol{a};\boldsymbol{b};A_+A_-)} \# \tag{2.43}$$

To ensure the decomposition of the unit operator, it is necessary to solve the following moment problem:

$$\int_0^\infty d\left(|z|^2\right) h_{red}\left(|z|^2\right)\left(|z|^2\right)^{s-1} == \frac{1}{\Gamma(a/b)}\rho(n) = \Gamma(s)\,\frac{\Gamma(s)\prod_{j=1}^{q}\Gamma\left(b_j-1+s\right)}{\prod_{i=1}^{p}\Gamma\left(a_i-1+s\right)} \tag{2.44}$$

which leads to the following expression of the integration measure:

$$d\mu(z) = \Gamma(a/b)\frac{d\varphi}{2\pi}d(|z|^2)\ {}_pF_q(\boldsymbol{a};\boldsymbol{b};|z|^2)\ G_{p,\,q+1}^{q+1,0}\left(|z|^2\left|\begin{matrix} /\,; & \boldsymbol{a-1} \\ 0\,,\boldsymbol{b-1}\,; & / \end{matrix}\right.\right) \tag{2.45}$$

Consequently, the resolution of identity leads to integral

$$\begin{aligned}\int \frac{d^2z}{\pi} G_{p,q+1}^{q+1,0}\left(|z|^2\left|\begin{matrix} /\,; & \boldsymbol{a-1} \\ 0,\boldsymbol{b-1}\,; & / \end{matrix}\right.\right) \#\,{}_pF_q(\boldsymbol{a};\boldsymbol{b};zA_+)\,{}_pF_q(\boldsymbol{a};\boldsymbol{b};z^*A_-)\# = \\ = \Gamma(b/a)\#\,{}_pF_q(\boldsymbol{a};\boldsymbol{b};A_+A_-)\#\end{aligned} \tag{2.46}$$

The angular integral is

$$\begin{aligned}\int_0^{2\pi}\frac{d\varphi}{2\pi}\#\,{}_pF_q(\boldsymbol{a};\boldsymbol{b};zA_+)\,{}_pF_q(\boldsymbol{a};\boldsymbol{b};z^*A_-)\# = \#\,{}_{2p}F_{2q+1}(\boldsymbol{a};\boldsymbol{a};1,\boldsymbol{b},\boldsymbol{b};A_+A_-|z|^2)\# = \\ = \left[\Gamma(b/a)\right]^2\#\,G_{2p,2q+2}^{1,2p}\left(-A_+A_-|z|^2\left|\begin{matrix} \boldsymbol{1-a}\,,\ 1-\boldsymbol{a}\,; & / \\ 0\,; & 0,\boldsymbol{1-b}\,,\ 1-\boldsymbol{b} \end{matrix}\right.\right)\#\end{aligned} \tag{2.47}$$

**Klauder-Perelomov nonlinear coherent states** (KP-NCSs) are defined by the action of a displacement operator on the vacuum Fock vector $|0>$, in the manner of Perelomov [Perelomov, 1972].

$$|\tilde{z}> = \frac{1}{\sqrt{\mathcal{N}\left(|\tilde{z}|^2\right)}}\exp\left(\tilde{z}A_+\right)|0> \tag{2.48}$$

and her development in the Fock – vector basis is

$$|\tilde{z}> = \frac{1}{\sqrt{\tilde{\mathcal{N}}\left(|\tilde{z}|^2\right)}}\sum_{n=0}^{\infty}\frac{\tilde{z}^n}{n!}\left(A_+\right)^n|0> = \frac{1}{\sqrt{\tilde{\mathcal{N}}\left(|\tilde{z}|^2\right)}}\sum_{n=0}^{\infty}\frac{\tilde{z}^n}{\sqrt{\tilde{\rho}(n)}}|n> \tag{2.49}$$

Due to the equality

$$(A_+)^n\,|\,0> = \sqrt{\rho(n)}\,|\,0> = \sqrt{n!}\,[f(n)]!\,|\,n> \tag{2.50}$$

it is observed that the connection between the structure functions of BG- and KP-NCSs is

$$\tilde{\rho}(n) = \frac{(n!)^2}{\rho(n)} = n!\frac{1}{\prod_{s=1}^{n}[f(s)]^2} = n!\frac{\prod_{i=1}^{p}(a_i)_n}{\prod_{j=1}^{q}(b_j)_n} \tag{2.51}$$

Then, the normalization function becomes

$$\begin{aligned}\tilde{\mathcal{N}}(|\tilde{z}|^2) &= \sum_{n=0}^{\infty}\frac{(|\tilde{z}|^2)^n}{\tilde{\rho}(n)} = \sum_{n=0}^{\infty}\frac{\prod_{j=1}^{q}(b_j)_n}{\prod_{i=1}^{p}(a_i)_n}\frac{(|\tilde{z}|^2)^n}{n!} = {}_q\tilde{F}_p(\boldsymbol{b}\,,\ \boldsymbol{a}\,;|\tilde{z}|^2) = \\ &= \Gamma(a/b)\tilde{G}^{1,q}_{q,p+1}\left(-|z|^2\left|\begin{matrix}\boldsymbol{1-b}\ ; & / \\ 0\ ; & \boldsymbol{1-a}\end{matrix}\right.\right)\end{aligned} \tag{2.52}$$

With this result, the expansion of KP-NCSs becomes

$$|\tilde{z}> = \frac{1}{\sqrt{{}_q\tilde{F}_p(\boldsymbol{b};\,\boldsymbol{a};|\tilde{z}|^2)}}\exp(\tilde{z}A_+)|\,0> = \frac{1}{\sqrt{{}_q\tilde{F}_p(\boldsymbol{b};\,\boldsymbol{a};|\tilde{z}|^2)}}\sum_{n=0}^{\infty}\frac{\tilde{z}^n}{\sqrt{\tilde{\rho}(n)}}\,|\,n> \tag{2.53}$$

The convergence radius $\tilde{R}_c$ of the KP-NCSs is calculated similarly:

$$\tilde{R}_c = \lim_{n\to\infty}\frac{\tilde{\rho}(n)}{\tilde{\rho}(n+1)} = \lim_{n\to\infty}\frac{(n!)^2}{\rho(n)}\frac{\rho(n+1)}{[(n+1)!]^2} = \lim_{n\to\infty}n^{q-p-1} = \begin{cases}\infty, & \text{if}\quad q-p-1>0\\ 1\,, & \text{if}\quad q-p-1>0\\ 0\,, & \text{if}\quad q-p-1>0\end{cases} \tag{2.54}$$

The inversion of the *p* and *q* indices can be observed, compared to the case of BG-NCSs.

The resolution of unity is

$$\begin{aligned}\hat{I} &= \int d\mu(\tilde{z})\,|\,\tilde{z}><\tilde{z}\,| = \\ &= \sum_{n,n'=0}^{\infty}\frac{|\,n><n'|}{\sqrt{\tilde{\rho}(n)}\sqrt{\tilde{\rho}(n')}}\int_0^{\infty}d(|\tilde{z}|^2)\frac{\tilde{h}(|\tilde{z}|^2)}{{}_q\tilde{F}_p(\boldsymbol{b};\,\boldsymbol{a};|\tilde{z}|^2)}\int_0^{2\pi}\frac{d\varphi}{2\pi}(\tilde{z}^*)^n(\tilde{z}')^{n'}\end{aligned} \tag{2.55}$$

The angular integral is $(|\tilde{z}|^2)\delta_{nn'}$ and with $\tilde{h}_{red}(|z|^2) \equiv \tilde{h}(|z|^2)\,/\,{}_q\tilde{F}_p(\boldsymbol{b};\,\boldsymbol{a};|\tilde{z}|^2)$ and $n = s-1$, the corresponding moment problem is

$$\int_0^\infty d(|\tilde{z}|^2)\,\tilde{h}_{red}(|\tilde{z}|^2)(|\tilde{z}|^2)^{s-1} = \tilde{\rho}(s-1) = \Gamma(b/a)\frac{\Gamma(s)\prod_{i=1}^{p}\Gamma(a_i - 1 + s)}{\prod_{j=1}^{q}\Gamma(b_j - 1 + s)} \tag{2.56}$$

whose solution is also a G-Meijer function, so that, finally, the integration measure becomes

$$d\mu(\tilde{z}) = \Gamma(b/a)\frac{d\varphi}{2\pi}d(|\tilde{z}|^2)\,\tilde{G}^{p+1,0}_{q,p+1}\left(|\tilde{z}|^2\,\middle|\,\begin{matrix} / \,; & \boldsymbol{b-1} \\ 0, \boldsymbol{a-1}\,; & / \end{matrix}\right)\,{}_q\tilde{F}_p(\boldsymbol{b};\,\boldsymbol{a};|\tilde{z}|^2) \tag{2.57}$$

Beginning from the unity decomposition relation, after a few simple operations we arrive at a new integral, in the real space of variable $|\tilde{z}|^2$, the validity of which can be verified also by using the properties of the G-Meijer functions:

$$\int_0^\infty d(|\tilde{z}|^2)\,\tilde{G}^{p+1,0}_{q,p+1}\left(|\tilde{z}|^2\,\middle|\,\begin{matrix} / \,; & \boldsymbol{b-1} \\ 0, \boldsymbol{a-1}\,; & / \end{matrix}\right)\exp\left(\tilde{z}^* A_+\right)\exp\left(\tilde{z}A_-\right) = \\ = \Gamma(a/b)\;{}_pF_q(\boldsymbol{a};\,\boldsymbol{b};\,A_+A_-) \tag{2.58}$$

We remind that, according to the DOOT rules, the operators can be treated as simple c-numbers, so they can be replaced with some arbitrary constants.

At the end of this section we will point out that the Meijer's G-function satisfies the following classical integral which we will use in the following [Mathai, 1973]:

$$\int_0^\infty dx\, x^{s-1}\, G^{m,n}_{p,q}\left(\beta x\,\middle|\,\begin{matrix} \{a_i\}_1^n\,; & \{a_i\}_{n+1}^p \\ \{b_j\}_1^m\,; & \{b_j\}_{m+1}^q \end{matrix}\right) = \frac{1}{\beta^s}\frac{\prod_{j=1}^{m}\Gamma(b_j + s)\prod_{i=1}^{n}\Gamma(1 - a_i - s)}{\prod_{j=m+1}^{q}\Gamma(1 - b_j - s)\prod_{i=n+1}^{p}\Gamma(a_i + s)} \tag{2.59}$$

## 3. Mehta's formula

Canonical CSs (associated with the one-dimensional harmonic oscillator, HO-1D) are obtained from the above formulas for the following particularization of the $p$ and $q$ indices: $p = q = 0$. It turns out that $\rho(n) = n!$ and ${}_0F_0\left(\;;\;;|z|^2\right) = \exp\left(|z|^2\right)$. Consequently, the canonical coherent states (CCSs) have the following development according to the Fock vectors:

$$|z\rangle = \exp\left(-\frac{1}{2}|z|^2\right)\sum_{n=0}^{\infty}\frac{z^n}{\sqrt{n!}}|n\rangle \tag{3.1}$$

They can also be obtained the *Q*- distribution function (Husimi's function)

$$Q(|z|^2) \equiv <z|\hat{\rho}|z> = \frac{1}{\bar{n}+1}\exp\left(-\frac{1}{\bar{n}+1}|z|^2\right) \tag{3.2}$$

as well as the *P*- quasi distribution function:

$$P(|z|^2) = \frac{1}{\bar{n}}\exp\left(-\frac{1}{\bar{n}}|z|^2\right) \tag{3.3}$$

Next, let's present some methods for obtaining the P-quasi distribution function, which are in certain stages equivalent, with many common considerations.

The ***Mehta's method.*** Using the anti-diagonal method, in 1967 Mehta [Mehta, 1967] derived a simple expression for the P-quasi distribution function for the case of pure and mixed (thermal) CCSs. Beginning from the diagonal representation of the density operator $\hat{\rho}$ the following equality is valid:

$$<-\alpha|\hat{\rho}|\alpha> = \int\frac{d^2z}{\pi}P(|z|^2)<-\alpha|z><z|\alpha> = e^{-|\alpha|^2}\int\frac{d^2z}{\pi}P(|z|^2)e^{-|z|^2}e^{z^*\alpha - z\alpha^*} \tag{3.4}$$

$$<-\alpha|\hat{\rho}|\alpha>e^{|\alpha|^2} = \int\frac{d^2z}{\pi}P(|z|^2)e^{-|z|^2}\exp\left(z^*\alpha - z\alpha^*\right) \tag{3.5}$$

Generally, the Fourier transform in complex plane is defined as

$$\mathcal{F}(\alpha) = \int\frac{d^2z}{\pi}f(z)\exp\left(z^*\alpha - z\alpha^*\right) \tag{3.6}$$

Consequently, the quantity $\mathcal{F}(\alpha) = <-\alpha|\hat{\rho}|\alpha>e^{|\alpha|^2}$ can be seen as the Fourier transform of function $f(z) = P(|z|^2)e^{-|z|^2}$ in the complex space.

Mehta showed that the following representation is valid, of course under the assumption that the integral exists:

$$f(z) = \frac{1}{\pi}\int\frac{d^2\alpha}{\pi}\mathcal{F}(\alpha)\exp\left(z\alpha^* - z^*\alpha\right) \tag{3.7}$$

i.e.

$$P(|z|^2)e^{-|z|^2} = \frac{1}{\pi}\int\frac{d^2\alpha}{\pi}e^{|\alpha|^2}<-\alpha|\hat{\rho}|\alpha>e^{z\alpha^* - z^*\alpha} \tag{3.8}$$

with $\alpha = \alpha_r + \mathrm{i}\alpha_i$ and $z = x + \mathrm{i}\,y$.

Let's briefly recall the steps of the demonstration of this formula. We considering a new complex variable $w = \alpha - \sigma = w_r + \mathrm{i}\, w_i$. Then, the function $f(z)$ can be written as

$$f(z) = \int \frac{d^2\sigma}{\pi^2} \mathcal{F}(\sigma) \exp\left(z\sigma^* - z^*\sigma\right) \tag{3.9}$$

Inserting this equation in Eq. (3.6), their right hand side (r.h.s.) successively becomes

$$\begin{aligned} \text{r.h.s.} &= \int \frac{d^2 z}{\pi} \left[ \int \frac{d^2\sigma}{\pi^2} \mathcal{F}(\sigma) \exp\left(z\sigma^* - z^*\sigma\right) \right] \exp\left(z^*\alpha - z\alpha^*\right) = \\ &= \int \frac{d^2\sigma}{\pi} \mathcal{F}(\sigma) \left[ \int \frac{d^2 z}{\pi^2} \exp\left(z\sigma^* - z^*\sigma\right) \exp\left(z^*\alpha - z\alpha^*\right) \right] \end{aligned} \tag{3.10}$$

The product of exponentials can be transformed as

$$\begin{aligned} \exp\left(z\sigma^* - z^*\sigma\right)\exp\left(z^*\alpha - z\alpha^*\right) = \exp\left[(\alpha-\sigma)z^* - (\alpha-\sigma)^* z\right] = \exp\left(w z^* - w^* z\right) = \\ = \exp\left(2\mathrm{i}\, w_i\, x\right)\exp\left(-2\mathrm{i}\, w_r\, y\right) \end{aligned} \tag{3.11}$$

and, because $\dfrac{d^2 z}{\pi^2} = \dfrac{dx}{\pi}\dfrac{dy}{\pi}$ we have

$$\begin{aligned} \text{r.h.s.} &== \int \frac{d^2\sigma}{\pi} \mathcal{F}(\sigma) \left[ \int \frac{d^2 z}{\pi^2} \exp\left(w z^* - w^* z\right) \right] = \\ &= \int \frac{d^2\sigma}{\pi} \mathcal{F}(\sigma) \left[ \frac{1}{2\pi} \int_{-\infty}^{+\infty} d(2x) \exp\left(2\mathrm{i}\, w_i\, x\right) \frac{1}{2\pi} \int_{-\infty}^{+\infty} d(2y)\, \exp\left(-2\mathrm{i}\, w_r\, y\right) \right] = \\ &= \int \frac{d^2\sigma}{\pi} \mathcal{F}(\sigma) \left[\delta(w_i)\,\delta(w_r)\right] = \int \frac{d^2\sigma}{\pi} \mathcal{F}(\sigma)\, \delta^2(w) = \int \frac{d^2\sigma}{\pi} \mathcal{F}(\sigma)\, \delta^2(\alpha - \sigma) = \mathcal{F}(\alpha) \end{aligned} \tag{3.12}$$

This shows that the relationship between the function $f(z) = P(|z|^2)\, e^{-|z|^2}$ and its Fourier transform in the complex space $\mathcal{F}(\alpha) = <-\alpha\,|\,\hat{\rho}\,|\,\alpha>e^{\,|\alpha|^2}$ is correct. This assertion will be the basis for deducing Mehta's generalized formula.

Pertinent analyzes of the validity conditions of Mehta's relationship were made later [Bishop, Vourdas, 1987], [Sounda, Mandal, 2022].

The ***integral method***. Next, we present another (may be, *less rigorous*, but much simpler) demonstration of Mehta's formula. Beginning from the relation (3.5)

$$<-\alpha\,|\,\hat{\rho}\,|\,\alpha>e^{\,|\alpha|^2} = \int \frac{d^2 z}{\pi} P(z)\, e^{-|z|^2} \exp\left(z^*\alpha - z\alpha^*\right) \tag{3.13}$$

let's multiply it by $\exp\left(\sigma\alpha^* - \sigma^*\alpha\right)$ and integrate after $\dfrac{d^2\alpha}{\pi}$ :

$$\begin{aligned}&\int \frac{d^2\alpha}{\pi} < -\alpha \,|\, \hat{\rho} \,|\, \alpha > e^{|\alpha|^2} \exp\left(\sigma\alpha^* - \sigma^*\alpha\right) = \\ &= \int \frac{d^2\alpha}{\pi} \exp\left(\sigma\alpha^* - \sigma^*\alpha\right) \int \frac{d^2 z}{\pi}\, e^{-|z|^2} e^{\alpha z^*} \left[e^{-\alpha^* z} P(z)\right]\end{aligned} \tag{3.14}$$

Using the mathematical formula [Renshan, 1990], [Popov et al, 2013]

$$\int \frac{d^2 z}{\pi}\, e^{-a|z|^2} e^{\sigma z^*} f(z) = \frac{1}{a} f\left(\frac{\sigma}{a}\right) \tag{3.15}$$

we obtain

$$\text{r.h.s.} = \int \frac{d^2\alpha}{\pi} e^{-|\alpha|^2} e^{\sigma\alpha^*} \left[e^{-\sigma^*\alpha} P(\alpha)\right] \tag{3.16}$$

and using one more time the formula (3.15), finally we obtain

$$P(\sigma) = e^{|\sigma|^2} \int \frac{d^2\alpha}{\pi} < -\alpha \,|\, \hat{\rho} \,|\, \alpha > e^{|\alpha|^2} \exp\left(\sigma\alpha^* - \sigma^*\alpha\right) \tag{3.17}$$

The ***characteristic function method***. At the end of this section we will point out that the expression of the *P – quasi distribution function* can be deduced also using the method of characteristic function [Rockower, 1988].

By definition, the characteristic function $C_N(\lambda, \lambda^*)$ is the expectation value of normally ordered product of exponential depending of canonical operators $a$ and $a^+$. Let's remember that the normal ordering of operators means that all creation operators are located on the left, and all annihilation operators are on the right of the respective expression. :

$$C_N(\lambda, \lambda^*) = <\# e^{\lambda a^+} e^{-\lambda a} \# > = \mathrm{Tr}\left(\# \rho\, e^{\lambda a^+} e^{-\lambda a} \#\right) \tag{3.18}$$

The partial derivatives of the characteristic function with respect to $\lambda$ and $\lambda^*$ generate the normally ordered moments of the creation and annihilation operators:

$$\left.\frac{\partial^n}{\partial\lambda^n} \frac{\partial^n}{\partial\lambda^{*n}} C_N(\lambda, \lambda^*)\right|_{\lambda=0} = (-1)^n <\# \left(a^+ a\right)^n \# > \tag{3.19}$$

Generally, if we have a function depending on the normally ordered product of creation and annihilation operators $\# f\left(a^{+}a\right)\# = \sum_{n} c_{n} \#\left(a^{+}a\right)^{n}\#$, then their expectation value in the CSs representation is

$$<\# f\left(a^{+}a\right)\#> = \int d\mu(z)\, P\left(|z|^{2}\right) f\left(|z|^{2}\right) \tag{3.20}$$

This relation is called the *optical equivalence theorem*, which implies the formal equivalence between expectation of normally ordered operators in quantum optics and the corresponding complex numbers in classical optics [Klauder, Streit, 1974].

As a consequence, we have

$$<\# e^{\lambda a^{+}} e^{-\lambda^{*} a}\#> = \int d\mu(z)\, P\left(|z|^{2}\right) \exp\left(\lambda z^{*} - \lambda^{*} z\right) \tag{3.21}$$

Calculating the two dimensional inversion Fourier transform we obtain the integral representation of the *P* quasi distribution function:

$$P\left(|z|^{2}\right) = \frac{1}{\pi}\int \frac{d^{2}\alpha}{\pi} \exp\left(z\alpha^{*} - z^{*}\alpha\right) \mathrm{Tr}\left\{\rho \exp\left(z a^{+} - z^{*} a\right)\right\} \tag{3.22}$$

## 4. Generalized Mehta's formula

Let's see now how to find the expression of the quasi-distribution P for the case of generalized or non-linear coherent states (NCSs). In other words, let's find Mehta's generalized formula. As we have seen in Eq. (2.42), the NCSs can be written in the following manner

$$|z> = \frac{1}{\sqrt{{}_{p}F_{q}(\boldsymbol{a}\,;\,\boldsymbol{b};|z|^{2})}}\, {}_{p}F_{q}(\boldsymbol{a}\,;\,\boldsymbol{b};|zA_{+})|0> \tag{4.1}$$

i.e. by generalization of the *normalized displaced operator for nonlinear CSs*.

$$\hat{\mathfrak{D}}(z) \equiv \frac{1}{\sqrt{{}_{p}F_{q}(\boldsymbol{a}\,;\,\boldsymbol{b};|z|^{2})}}\, {}_{p}F_{q}(\boldsymbol{a}\,;\,\boldsymbol{b};|zA_{+})\;, \qquad |z> = \hat{\mathfrak{D}}(z)|0> \tag{4.2}$$

This generalized displaced operator acting on the vacuum state $|0>$ was first introduced by Mojaveri and Dehghani, for the case of pseudoharmonic oscillator (PHO) [Mojaveri, Dehghani, 2013]. Due to the equality

$${}_{p}F_{q}(\boldsymbol{a}\,;\,\boldsymbol{b};|z^{*}A_{-})|0> = |0> \tag{4.3}$$

we can also write an equivalent expression

$$\hat{\mathfrak{D}}(z) \equiv \frac{1}{\sqrt{{}_pF_q(\boldsymbol{a}\,;\,\boldsymbol{b};|\,z\,|^2)}}\,{}_pF_q(\boldsymbol{a}\,;\,\boldsymbol{b};|\,zA_+)\ {}_pF_q(\boldsymbol{a}\,;\,\boldsymbol{b};|\,z^*A_-)\ , \tag{4.4}$$

but we will not use this way of writing.

This last form of writing is the analogue of the "orthodox" normalized displaced operator for the canonical CSs $\hat{\mathfrak{D}}(z) = \exp(-|\,z\,|^2)\exp\left(z\,a^+ - z^* a\right)$ .

For a function that depends on product of the generalized creation and annihilation operators, in DOOT acceptance, $\#\, f\left(A_+A_-\right)\# = \sum_n c_n \#\left(A_+A_-\right)^n \#$, some matrix element in the NCSs representation is

$$<\alpha\,|\#\, f\left(A_+A_-\right)\#|\,z> = \sum_n c_n <\alpha|\#\left(A_+A_-\right)^n\#|\,z> = f\left(\alpha^* z\right), \tag{4.5}$$

and the generalized *optical equivalence theorem* reads

$$<\#\, f\left(A_+A_-\right)\#> = \mathrm{Tr}\left[\rho\#\, f\left(A_+A_-\right)\#\right] = \int d\mu(z)\, P\left(|\,z\,|^2\right) f\left(|\,z\,|^2\right) \tag{4.6}$$

Generally, if we consider the Hamiltonian eigenequation as

$$\hat{\mathcal{H}}\,|\,n> = E_n\,|\,n> = \hbar\,\omega\, e(n)\,|\,n> \tag{4.7}$$

the density operator can be written, on the one hand, as

$$\begin{aligned}\rho &= \frac{1}{Z(\beta_T)}\sum_n \exp\left(-\beta_T\, E_n\right)|\,n><n\,| = \\ &= \frac{1}{\#\,{}_pF_q(\boldsymbol{a}\,;\,\boldsymbol{b};A_+A_-)\,\#}\sum_n \frac{1}{Z(\beta_T)}\left(\frac{\bar{n}}{\bar{n}+1}\right)^{e(n)} \frac{\#\left(A_+A_-\right)^n\ \#}{\rho(n)}\end{aligned} \tag{4.8}$$

and on the other hand, like

$$\begin{aligned}\rho &= \int d\mu(z) P\left(|\,z\,|^2\right) z><z\,| = \\ &= \frac{1}{\#\,{}_pF_q(\boldsymbol{a}\,;\,\boldsymbol{b};A_+A_-)\,\#}\int d\mu(z) P\left(|\,z\,|^2\right)\frac{1}{{}_pF_q(\boldsymbol{a}\,;\,\boldsymbol{b};|\,z\,|^2)}\#\,{}_pF_q(\boldsymbol{a}\,;\,\boldsymbol{b};zA_+)\,{}_pF_q(\boldsymbol{a}\,;\,\boldsymbol{b};z^*A_-)\,\#\end{aligned} \tag{4.9}$$

The last equation becomes, after angular integration

$$\begin{aligned}\rho &= \frac{1}{\#\,{}_pF_q(\boldsymbol{a}\,;\,\boldsymbol{b};A_+A_-)\,\#}\times \\ &\times \sum_n \frac{\#\left(A_+A_-\right)^n\ \#}{\left[\rho(n)\right]^2}\left[\Gamma\left(a/b\right)\int_0^\infty d\left(|\,z\,|^2\right) G_{p,q+1}^{q+1,0}\left(|\,z\,|^2 \,\middle|\, \begin{matrix} / \,; & \boldsymbol{a-1} \\ 0,\boldsymbol{b-1}\,; & / \end{matrix}\right) P\left(|\,z\,|^2\right)\left(|\,z\,|^2\right)^n\right]\end{aligned} \tag{4.10}$$

For the two expressions of $\rho$ to be equal, it is obvious that the integral in the right bracket must fulfill the condition:

$$\int_0^\infty d\left(|z|^2\right) G_{p,q+1}^{q+1,0}\left(|z|^2 \left| \begin{array}{cc} / \,; & \mathbf{a}-\mathbf{1} \\ 0,\mathbf{b}-\mathbf{1} \,; & / \end{array} \right.\right) P\left(|z|^2\right)\left(|z|^2\right)^n = \frac{1}{Z(\beta_T)}\left(\frac{\bar{n}}{\bar{n}+1}\right)^{e(n)} \frac{1}{\Gamma\left(a/b\right)}\rho(n) \qquad (4.11)$$

Next, the calculation depends on the concrete expression of the dimensionless eigenvalues $e(n)$. For linear energy spectrum we have $e(n) = n + e_0$. Substituting it in the above condition we will arrive at the following expression for *P* quasi distribution:

$$P\left(|z|^2\right) = \frac{1}{\bar{n}} \frac{G_{p,q+1}^{q+1,0}\left(\frac{\bar{n}+1}{\bar{n}}|z|^2 \left| \begin{array}{cc} / \,; & \mathbf{a}-\mathbf{1} \\ 0,\mathbf{b}-\mathbf{1} \,; & / \end{array} \right.\right)}{G_{p,q+1}^{q+1,0}\left(|z|^2 \left| \begin{array}{cc} / \,; & \mathbf{a}-\mathbf{1} \\ 0,\mathbf{b}-\mathbf{1} \,; & / \end{array} \right.\right)} \qquad (4.12)$$

If we calculate the diagonal elements of the density operator $\rho$, i.e. Husimi's function $<\alpha|\rho|\alpha>= Q(\alpha)$, we will have

$$<\alpha|\rho|\alpha>= \int d\mu(z)\, P\left(|z|^2\right) <\alpha|z><z|\alpha> \qquad (4.13)$$

or further

$$Q(\alpha) \equiv <\alpha|\hat{\rho}|\alpha>= \frac{1}{\bar{n}+1} \frac{{}_pF_q(\mathbf{a}\,;\mathbf{b};\frac{\bar{n}}{\bar{n}+1}|\alpha|^2)}{{}_pF_q(\mathbf{a}\,;\mathbf{b};|\alpha|^2)} \qquad (4.14)$$

For the case of quantum systems with a linear spectrum, expressed in terms of operators, Eq. (4.7) becomes

$$\begin{gathered} \frac{1}{\bar{n}+1} \#\, {}_pF_q(\mathbf{a}\,;\mathbf{b};\frac{\bar{n}}{\bar{n}+1}A_+A_-)\# = \Gamma\left(a/b\right)\times \\ \times\int_0^\infty d\left(|z|^2\right) G_{p,q+1}^{q+1,0}\left(|z|^2 \left| \begin{array}{cc} / \,; & \mathbf{a}-\mathbf{1} \\ 0,\mathbf{b}-\mathbf{1} \,; & / \end{array} \right.\right) P\left(|z|^2\right) \int_0^{2\pi}\frac{d\varphi}{2\pi} \#\, {}_pF_q(\mathbf{a}\,;\mathbf{b};zA_+)\, {}_pF_q(\mathbf{a}\,;\mathbf{b};z^*A_-)\# \end{gathered} \qquad (4.15)$$

After performing the angular integration, the r.h.s. becomes

$$\text{r.h.s.} = \sum_n \frac{\#\left(A_+A_-\right)^n\#}{\left[\rho(n)\right]^2} \int_0^\infty d\left(|z|^2\right) G_{p,q+1}^{q+1,0}\left(|z|^2 \left| \begin{array}{cc} / \,; & \mathbf{a}-\mathbf{1} \\ 0,\mathbf{b}-\mathbf{1} \,; & / \end{array} \right.\right) P\left(|z|^2\right)\left(|z|^2\right)^n \qquad (4.16)$$

After the change of function $\widehat{P}\left(|z|^2\right) \equiv G_{p,q+1}^{q+1,0}\left(|z|^2 \left| \begin{matrix} / \; ; & \boldsymbol{a-1} \\ 0, \boldsymbol{b-1} \; ; & / \end{matrix} \right.\right) P\left(|z|^2\right)$ it is observed that the integral must have the value

$$\int_0^\infty d\left(|z|^2\right) \widehat{P}\left(|z|^2\right) \left(|z|^2\right)^n = \frac{1}{\Gamma(a/b)} \left(\frac{\bar{n}}{\bar{n}+1}\right)^{n+1} \rho(n) \; . \tag{4.17}$$

This leads to the solution (4.9).

Therefore, in Eq. (4.10), $<\alpha|z><z|\alpha>$ play the role of the normalized generalized displacement operator, and $<\alpha|\rho|\alpha>$ is the generalized Fourier transform of the function $P\left(|z|^2\right)$ in the complex space.

More clearly, moving to complex variables, Eq. (4.10) will be written in the form

$$\begin{gathered} <\alpha|\hat{\rho}|\alpha> {}_pF_q(\boldsymbol{a}\,;\boldsymbol{b};|\alpha|^2) = \\ \Gamma\left(a/b\right) \int \frac{d^2 z}{\pi} G_{p,q+1}^{q+1,0}\left(|z|^2 \left| \begin{matrix} / \; ; & \boldsymbol{a-1} \\ 0, \boldsymbol{b-1} \; ; & / \end{matrix} \right.\right) P\left(|z|^2\right) {}_pF_q(\boldsymbol{a}\,;\boldsymbol{b};z\alpha^*) {}_pF_q(\boldsymbol{a}\,;\boldsymbol{b};z^*\alpha) \end{gathered} \tag{4.18}$$

from where the role played by the general normalized displacement operators can be better seen.

Let's see if the generalized inverse Fourier transform is also valid, that is, if $P\left(|\alpha|^2\right)$ is the generalized inverse transform of $<z|\rho|z>$.

$$P\left(|\alpha|^2\right) = \int d\mu(z) <z|\rho|z><\alpha|z><z|\alpha> \tag{4.19}$$

Substituting the expression for $<z|\rho|z>$ and performing similar calculations, we will arrive at the expression (4.9), which proves that the above statement is correct.

Let's highlight that all the calculations above were made for NCSs defined in the Barut-Girardello manner. We have the task of verifying whether the generalized Fourier transform and its inverse is also applicable for NCSs defined in the Klauder-Perelomov manner.

From definition, the Klauder-Perelomov nonlinear coherent states (KP-NCSs) are

$$|\tilde{z}> = \frac{1}{\sqrt{{}_q\widetilde{F}_p(\boldsymbol{b}\,;\boldsymbol{a};|\tilde{z}|^2)}} \exp\left(\tilde{z}A_+\right)|0> \tag{4.20}$$

which shows that the normalized generalized displacement operator for these states is

$$\hat{\tilde{\mathcal{D}}}(\tilde{z}) \equiv \frac{1}{\sqrt{{}_q\tilde{F}_p(\boldsymbol{b}\,;\boldsymbol{a};|\,\tilde{z}\,|^2)}}\exp\left(\tilde{z}A_+\right) \;, \qquad |\,\tilde{z}> = \hat{\tilde{\mathcal{D}}}(\tilde{z})\,|\,0> \tag{4.21}$$

If we replace in the expression of the diagonal development of the density operator and take into account the DOOT rules, we get

$$\rho = \frac{1}{\#_q\tilde{F}_p(\boldsymbol{b}\,;\boldsymbol{a};A_+A_-)\#}\int d\mu(\tilde{z})\,\tilde{P}\left(|\,\tilde{z}\,|^2\right)\frac{\#\exp\left(\tilde{z}A_+\right)\exp\left(\tilde{z}^*A_-\right)\#}{{}_q\tilde{F}_p(\boldsymbol{b}\,;\boldsymbol{a};|\,\tilde{z}\,|^2)} \tag{4.22}$$

Taking into account the relation,

$$\exp\left(\tilde{z}^*A_-\right)|\,\tilde{\alpha}> = \exp\left(\tilde{z}^*\tilde{\alpha}\right) \tag{4.23}$$

as well as its complex conjugate, if we calculate the diagonal elements of $\rho$, we will obtain:

$$\begin{aligned} &<\tilde{\alpha}\,|\,\hat{\rho}\,|\,\tilde{\alpha}>\,{}_qF_p(\boldsymbol{b}\,;\,\boldsymbol{a};|\,\tilde{\alpha}\,|^2) = \\ &= \Gamma\left(a/b\right)\int\frac{d^2z}{\pi}\tilde{G}^{p+1,0}_{q,p+1}\left(|\,z\,|^2\left|\begin{array}{cc} /\;; & \boldsymbol{b}-\boldsymbol{1} \\ 0,\,\boldsymbol{a}-\boldsymbol{1}\;; & / \end{array}\right.\right)\tilde{P}\left(|\,\tilde{z}\,|^2\right)\exp\left(\tilde{z}\,\tilde{\alpha}^*\right)\exp\left(\tilde{z}^*\tilde{\alpha}\right) \end{aligned} \tag{4.24}$$

This shows that $<\tilde{\alpha}\,|\,\hat{\rho}\,|\,\tilde{\alpha}>\,{}_qF_p(\boldsymbol{b}\,;\,\boldsymbol{a};|\,\tilde{\alpha}\,|^2)$ is the generalized Fourier transform of the complex space of the function $P\left(|\,\tilde{z}\,|^2\right)$ in the case of KP-NCSs.

The reciprocal transformation is also valid, that is

$$\tilde{P}\left(|\,\tilde{z}\,|^2\right) = \int d\tilde{\mu}(\tilde{\alpha}) <\tilde{\alpha}\,|\,\rho\,|\,\tilde{\alpha}><\tilde{z}\,|\,\tilde{\alpha}><\tilde{\alpha}\,|\,\tilde{z}> \tag{4.25}$$

which can be verified by similar calculations.

In conclusion, the generalized Fourier transform in the complex space of the function $f\left(z\right)$ in the case of NCSs is, generally

$$\mathcal{F}(\alpha) = \int d\mu(z)\; f\left(z\right) <\alpha|\,z><z\,|\,\alpha> \tag{4.26}$$

as well as the corresponding inverse transform

$$f(z) = \int d\mu(\alpha)\,\mathcal{F}(\alpha) < z\,|\,\alpha><\alpha\,|\,z> \tag{4.27}$$

These last two equations, as new obtained results, are counterparts of the Fourier transform for canonical CSs, Eqs. (3.6) and (3.7), but applicable to arbitrary NCSs.

## 5. An example

Let's illustrate the results obtained above with an example. We will examine the case of the pseudoharmonic oscillator (PHO), whose effective potential is [Popov, JPA-MG, 2001]:

$$V_J(r) = \frac{m_{red}\,\omega^2}{8} r_0^2 \left( \frac{r}{r_0} - \frac{r_0}{r} \right)^2 + \frac{\hbar^2}{2m_{red}} J(J+1) \frac{1}{r^2} \tag{5.1}$$

where $m_{red}$ and $\omega$ are, respectively, the reduced mass and the angular frequency of the PHO, $r_0$ is the equilibrium distance and $J = 0, 1, 2, \ldots$ is the rotational quantum number. A suitable physical system which can be correctly modeled by the PHO is the diatomic molecule, e.g.

The time independent Schrödinger equation for PHO is

$$H\left| n; k \right> = E_{nJ} \left| n; k \right> \tag{5.2}$$

where the energy eigenvalues (the rotational – vibrational energy spectrum) are

$$E_{nJ} = \hbar\omega\left( n + \frac{1}{2} \right) + \frac{\hbar\omega}{2}(2k-1) - \frac{m_{red}\omega^2}{4} r_0^2 \equiv \hbar\omega(n+k) - \frac{m_{red}\omega^2}{4} r_0^2 \tag{5.3}$$

Here the rotational quantum number $J$ is embedded in the Bargmann index $k = k(J)$, labeling the irreducible unitary representations of the associated quantum group. Their expression is

$$k = \frac{1}{2} + \frac{1}{2}\sqrt{\left( J + \frac{1}{2} \right)^2 + \left( \frac{m_{red}\omega}{2\hbar} r_0^2 \right)^2} \,. \tag{5.4}$$

If we neglect the last term, the dimensionless energy expression

$$e_k(n) = n + k \tag{5.5}$$

i.e. the energy spectrum of PHO is linear regarding to the main quantum number $n$.

Consequently, for PHO we have the following characteristics: $p = 1$ , $a_1 = 1$, $q = 1$ and $b_1 = k + 1$, the structure function is $\rho(n) = (k+1)_n$ . In short: $\left| n; k \right> \equiv \left| n \right>$.

$$A_- \,|\, n> = \sqrt{e_k(n)} \,|\, n-1> \quad , \quad A_+ \,|\, n> = \sqrt{e_k(n+1)} \,|\, n+1> \quad , \quad A_+ A_- \,|\, n> = e_k(n) \,|\, n> \tag{5.6}$$

Consequently, the ***BG-NCSs*** is defined as

$$|z>=\frac{1}{\sqrt{{}_1F_1(1;k+1;|z|^2)}}\sum_{n=0}^{\infty}\frac{z^n}{\sqrt{(k+1)_n}}|n>= \\ =\frac{1}{\sqrt{{}_1F_1(1;k+1;|z|^2)}}{}_1F_1(1;k+1;|zA_+)|0> \tag{5.7}$$

so the normalized generalized displacement operator is

$$\hat{\mathfrak{D}}(z)=\frac{1}{\sqrt{{}_1F_1(1;k+1;|z|^2)}}{}_1F_1(1;k+1;|zA_+) \tag{5.8}$$

From the decomposition of unity operator it follows that the integration measure have the expression

$$d\mu(z)=\frac{1}{\Gamma(k+1)}\frac{d^2z}{\pi}\exp\left(-|z|^2\right)\left(|z|^2\right)^k{}_1F_1(1;k+1;|z|^2) \tag{5.9}$$

Following the rules from the previous sections, the following expressions for the BG-NCSs are easily reached [Popov, 2017]:

- the normalized density operator

$$\hat{\rho}=\frac{1}{\bar{n}+1}\#\frac{{}_1F_1(1;k+1;\frac{\bar{n}}{\bar{n}+1}A_+A_-)}{{}_1F_1(1;k+1;A_+A_-)}\# \tag{5.10}$$

- the Husimi's function

$$Q(z)\equiv<z|\hat{\rho}|z>=\frac{1}{\bar{n}+1}\frac{{}_1F_1(1;k+1;\frac{\bar{n}}{\bar{n}+1}|z|^2)}{{}_1F_1(1;k+1;|z|^2)} \tag{5.11}$$

- the *P*-quasi distribution function

$$P(|z|^2)=\frac{1}{\bar{n}}\left(\frac{\bar{n}+1}{\bar{n}}\right)^k\exp\left(-\frac{1}{\bar{n}}|z|^2\right) \tag{5.12}$$

Let's check if $<z|\hat{\rho}|z>{}_1F_1(1;k+1;|z|^2)$ is the generalized Fourier transform from the cpmplex space of the *P*-quasi distribution function $P(|z|^2)$.

Starting from the diagonal form of the density operator, it can also be written in the form

$$\hat{\rho} = \frac{1}{\#_1F_1(1;k+1;A_+A_-)\#}\int d\mu(z)\,P(|z|^2)\frac{\#\,{}_1F_1(1;k+1;zA_+)\,{}_1F_1(1;k+1;z^*A_-)\#}{{}_1F_1(1;k+1;|z|^2)} \tag{5.13}$$

such that its diagonal elements are

$$<\alpha|\hat{\rho}|\alpha>= \frac{1}{\#_1F_1(1;k+1;|\alpha|^2)\#}\int d\mu(z)\,P(|z|^2)\frac{\#\,{}_1F_1(1;k+1;z\alpha^*)\,{}_1F_1(1;k+1;z^*\alpha)\#}{{}_1F_1(1;k+1;|z|^2)} \tag{5.14}$$

After performing the angular integral and replacing $P(|z|^2)$ with its expression above (5.12), we will obtain the correct expression (5.11) for the diagonal elements of $\hat{\rho}$, so implicitly the previous statement is correct. The validity of the inverse Fourier transform is demonstrated analogously.

Let's do the same for KP-NCSs.

The ***KP-NCSs*** is defined as

$$|\tilde{z}>= \frac{1}{\sqrt{{}_1\tilde{F}_1(k+1;1;|\tilde{z}|^2)}}\exp\left(\tilde{z}A_+\right)|0>= \frac{1}{\sqrt{{}_1\tilde{F}_1(k+1;1;|\tilde{z}|^2)}}\sum_{n=0}^{\infty}\sqrt{\frac{(k+1)_n}{[(1)_n]^2}}\tilde{z}^n|n> \tag{5.15}$$

and the corresponding normalized generalized displacement operator is

$$\hat{\tilde{\mathcal{D}}}(z) = \frac{1}{\sqrt{{}_1\tilde{F}_1(k+1;1;|\tilde{z}|^2)}}\exp\left(\tilde{z}A_+\right) \tag{5.16}$$

From the decomposition of unity operator we obtain that the integration measure is

$$d\mu(\tilde{z}) = \Gamma(k+1)\frac{d\varphi}{2\pi}d(|\tilde{z}|^2)\,\tilde{G}^{2,0}_{1,2}\left(|\tilde{z}|^2\left|\begin{matrix} /\,; & k \\ 0,\,0\,; & / \end{matrix}\right.\right){}_1\tilde{F}_1(k+1;1;|\tilde{z}|^2) \tag{5.17}$$

For the KP-NCSs the following expressions are valid:

- the normalized density operator expressed in terms of the DOOT ordered product of operators $A_+A_-$ is, naturally the same as for BG-NCSs

$$\hat{\rho} = \frac{1}{\bar{n}+1}\#\frac{{}_1F_1(1;k+1;\frac{\bar{n}}{\bar{n}+1}A_+A_-)}{{}_1F_1(1;k+1;A_+A_-)}\# \tag{5.18}$$

On the other hand, it is developed as

$$\begin{aligned}\rho &= \int d\mu(\tilde{z})\tilde{P}\left(|\tilde{z}|^2\right)\tilde{z}><\tilde{z}| = \Gamma(k+1)\sum_n|n><n|\frac{(k+1)_n}{[(1)_n]^2}\times \\ &\times\int_0^\infty d(|\tilde{z}|^2)\frac{d\varphi}{2\pi}\,\tilde{G}^{2,0}_{1,2}\left(|\tilde{z}|^2\left|\begin{matrix} /\,; & k \\ 0,\,0\,; & / \end{matrix}\right.\right)\tilde{P}(|\tilde{z}|^2)(|\tilde{z}|^2)^n\end{aligned} \tag{5.19}$$

- as well as the Husimi's function (is the same as for BG-NCSs, but depending on $\tilde{\alpha}$ )

$$\tilde{Q}(\tilde{\alpha}) \equiv < \tilde{\alpha} \,|\, \hat{\rho} \,|\, \tilde{\alpha} > = \frac{1}{\bar{n}+1} \frac{{}_1F_1(1; k+1; \frac{\bar{n}}{\bar{n}+1} |\, \tilde{\alpha} \,|^2 )}{{}_1F_1(1; k+1; |\, \tilde{\alpha} \,|^2 )} \tag{5.20}$$

- the *P*-quasi distribution function

$$\tilde{P}(|\, \tilde{z} \,|^2) = \frac{1}{\bar{n}} \frac{\tilde{G}_{1,2}^{2,0}\left( \frac{\bar{n}+1}{\bar{n}} |\, \tilde{z} \,|^2 \middle| \begin{array}{cc} / \,; & k \\ 0,\, 0\,; & / \end{array} \right)}{\tilde{G}_{1,2}^{2,0}\left( |\, \tilde{z} \,|^2 \middle| \begin{array}{cc} / \,; & k \\ 0,\, 0\,; & / \end{array} \right)} \tag{5.21}$$

Let's check if $< \tilde{\alpha} \,|\, \hat{\rho} \,|\, \tilde{\alpha} > {}_1F_1(1; k+1; |\, \tilde{\alpha} \,|^2 )$ is the generalized Fourier transform from the cpmplex space of the *P*-quasi distribution function $P(|\, z \,|^2)$.

The diagonal form of the density operator, it can also be written as well

$$\hat{\rho} = \frac{1}{\#{}_1\tilde{F}_1(k+1; 1\,; A_+ A_-)\#} \int d\mu(\tilde{z})\, \tilde{P}(|\, \tilde{z} \,|^2) \frac{\#\,{}_1\tilde{F}_1(k+1; 1\,; \tilde{z} A_+ )\; {}_1\tilde{F}_1(k+1; 1; \tilde{z}^* A_- )\#}{{}_1\tilde{F}_1(k+1; 1\,; |\, \tilde{z} \,|^2 )} \tag{5.22}$$

such that taking into consideration its diagonal elements, we have

$$\begin{aligned} &< \tilde{\alpha} \,|\, \hat{\rho} \,|\, \tilde{\alpha} > {}_1F_1(k+1; 1\,; |\, \tilde{\alpha} \,|^2 ) = \\ &= \Gamma(k+1) \int_0^\infty d(|\, \tilde{z} \,|^2)\; \tilde{G}_{1,2}^{2,0}\left( |\, \tilde{z} \,|^2 \middle| \begin{array}{cc} / \,; & k \\ 0,\, 0\,; & / \end{array} \right) \tilde{P}(|\, \tilde{z} \,|^2) \int_0^{2\pi} \frac{d\varphi}{2\pi} {}_1\tilde{F}_1(k+1; 1\,; \tilde{z}\tilde{\alpha}^* )\; {}_1\tilde{F}_1(k+1; 1\,; \tilde{z}^* \tilde{\alpha} ) \end{aligned} \tag{5.23}$$

After performing the angular integration

$$\int_0^{2\pi} \frac{d\varphi}{2\pi} {}_1\tilde{F}_1(k+1; 1\,; \tilde{z}\tilde{\alpha}^* )\; {}_1\tilde{F}_1(k+1; 1\,; \tilde{z}^* \tilde{\alpha} ) = \sum_n \frac{(k+1)_n}{\left[(1)_n\right]^2} (|\, \tilde{\alpha} \,|^2)^n \, (|\, \tilde{z} \,|^2)^n \tag{5.24}$$

and replacing $\tilde{P}(|\, \tilde{z} \,|^2)$ with its expression above (5.21), we will obtain the correct expression (5.18) for the diagonal elements of $\hat{\rho}$, which demonstrate that the normalized generalized inverse Fourier transform in complex space is also correct.

At the end of this section we want to mention that for the particular case $k = 0$, all these expression becomes identical with those for the one dimensional harmonic oscillator (HO-1D).

## 6. Concluding remarks

The derivation of the expression *P*-quasi distribution function as the inverse Fourier transform of Husimi's function (the diagonal elements of the density operator in the representation of coherent states) has been a target almost since the beginning of the discovery of coherent states. Even two of the pioneers of the formalism of coherent states (Glauber and Sudarshan) posed this problem, aiming at the interpretation of the quasi-distribution function *P*- that appears in the diagonal development of the density operator with respect to the projectors of coherent states. For the canonical coherent states, considered to be linear, associated with the one-dimensional harmonic oscillator, the problem was solved in 1967 by Mehta. The same problem remained unsolved for the general case, that is, for a certain nonlinear coherent state. Our paper set out to solve this problem.

We showed that the Fourier transform for the complex space can be generalized for a certain nonlinear coherent state. Applied to the pair of distributions, Husimi's $<\alpha|\hat{\rho}|\alpha>\mathcal{N}(|\alpha|^2)$ and *P*-quasi distribution $P\left(|z|^2\right)$, the relationship between them is maintained as in the case of canonical coherent states. From the calculations it turned out that Husimi's (unnormed) function $<\alpha|\hat{\rho}|\alpha>\mathcal{N}(|\alpha|^2)$ is the generalized Fourier transform for the complex space of the quasi-distribution $P\left(|z|^2\right)$, and vice versa. The role of the normed displacement operator is played by the product $<\alpha|z><z|\alpha>$. By applying it to a concrete case, that of the pseudoharmonic oscillator, we illustrated the correctness of the general theoretical results obtained. The Fourier transform for the complex space, versus the pair of distribution functions (*Q* and *P*), is valid for both types of nonlinear coherent states, Barut-Girardello and Klauder-Perelomov, respectively.